\documentclass{elsart}
\usepackage{amsmath}
\usepackage{amssymb}
\usepackage{epsfig}
\usepackage{float}   
\begin{document}
\runauthor{Calzavarini, Schifano, Tripiccione and Vicer\'e}
\begin{frontmatter}
\title{ Matched filters for coalescing binaries detection on
massively parallel computers}

\author[ferrara]{E. Calzavarini\thanksref{mail}}, 
\author[pisa]{L. Sartori},
\author[pisa]{F. Schifano},
\author[ferrara]{R. Tripiccione},
\author[urbino]{A. Vicer\'e}
\address[ferrara]{Physics Department, Universit\`a di Ferrara and\\ INFN, Sezione di Ferrara I-44100, Ferrara, Italy}
\address[pisa]{ INFN, Sezione di Pisa, I-56010 S.Piero a Grado (PI), Italy} 
\address[urbino]{Physics Department, Universit\`a di Urbino I-61029 Urbino, Italy } 
\thanks[mail]{{\it Corresponding author:}  calzavar@fe.infn.it ; Dipartimento di Fisica, via Paradiso 12, I-44100 Ferrara, Italy. } 
                                                          
\begin{abstract}
In this paper we discuss some computational problems associated to matched
filtering of experimental signals from gravitational wave interferometric
detectors in a parallel-processing environment. We then specialize
our discussion to the use of the APEmille and apeNEXT processors for this task.
Finally, we accurately estimate the performance of an APEmille system on
a computational load appropriate for the LIGO and VIRGO experiments, and
extrapolate our results to apeNEXT.
\end{abstract}
\end{frontmatter}

\section{Introduction}

Several earth-based interferometric experiments for the detection of
gravitational waves (GW) are currently under development, and expected to reach
the data-taking stage in the near future. On a longer time scale, space-based
experiments are foreseen \cite{giazotto}. These experiments will search, among
other, for GW generated by inspiralling compact binary-star systems.

The expected functional form of the signal produced by a coalescing system is
known to good approximation \cite{blanchet}, so matched filtering is an effective
strategy to extract GW signals from the noise background. Matched filtering is
basically obtained by projecting the experimental output (signal plus noise) onto
the expected theoretical signal, and is best done in Fourier space, using Fast
Fourier Transform (FFT) techniques (see later for more details). The functional
form of the expected signal depends however on the physical parameters (e.g.,
masses, angular momenta, eccentricity) of the inspiralling system. It is necessary
to match the experimental output to a set of expected signals (so-called
templates) corresponding to points in the parameter space that cover the physical
region of interest and are close enough (under some appropriate metric) to ensure
sufficient overlap with any expected GW event.

The number of needed templates for e.g. the VIRGO experiment is of
order of $10^5 \cdots 10^6$, so the corresponding computational cost is huge
by current standards. 
A nice requirement is the possibility of real-time analysis of the
experimental data, which means that the available computational power is enough
to process experimental data at the rate at which they are produced, so
a prompt ``trigger'' of a GW event is possible \cite{ligoexp}. 

Matched filtering to a (large) set of templates is an obvious candidate for
parallel processing of the simplest form, e.g., data farming with all elements
of the farm performing the same computation (Single Program Multiple Data
(SPMD) processing). Indeed, the experimental data stream is sent to all
processors in the farm, each element performing the matching procedures for a subset of the physical templates.
 
Massively parallel specialized SPMD architectures, with peak processing power
of the order of 1 Tflops have been developed by several groups to fulfill the
computational requirements of Lattice Gauge Theories (LGT) \cite{norman}.  In this paper we
want to analyze the performance of one such system (the APEmille system \cite{ape}) for matched filtering of GW signals.

This paper is not a proposal to use APE-like
systems in an actual experimental (the relative merits of different computer
systems in a large experiment have so many facets that they can only be
assessed by those directly working on it).
Rather, the potential usefulness of our work
lies in the following: given the fast pace of development in the computer
industry, an experiment will try to delay the commissioning of a production
system to as late a point in time as possible, since huge gains in price and/or
price/performance can be expected.  This means that very large computing
capabilities will not be available for much needed early tests and simulations.
APE systems might provide an answer to this problem.

The focus of this paper is the measurement of the performance of APE systems
for matched filtering. Some parts of the paper have however a more general
scope and refer to general parallelization criteria for the problem at hand.

This paper is structured as follows: Section 2 briefly reviews the formalism of
matched filtering. Section 3 evaluates the associated computational cost in
general terms and discusses some strategies to minimize this quantity. Section 4
discusses the features of the APE systems relevant for the problem, while section
5 presents a procedure for allocation of templates to processors suitable for APE
and general enough to adapt to other computer systems. Section 6 presents the
result of actual performance measurements made on APE, while section 7 contains
our concluding remarks.

\section{Formalism}

In this section we briefly summarize the mathematical formalism recently developed
to analyze matched filtering of GW signals from  coalescing binaries. We closely follow the notation presented in \cite{owen}.   

We call $h(t)$ the interferometer output, which is the sum of the signal $s(t)$
and the noise $n(t)$, while $u(t)$ is a template. $n(t)$ is characterized by its
one-sided spectral density:    
\begin{equation}
E[ \tilde{n}(f_{1}) \ \tilde{n}^{\ast}(f_{2}) ] = \frac{1}{2}\  \delta(f_{1} - f_{2})\ S_{n}(|f_{1}|)
\end{equation}
where $E[\dots ]$ means ensemble expectation value, tilde ($\sim$) stands for
Fourier transformed functions and asterisk ($\ast$) for complex
conjugation.

For the sake of definiteness, we consider in the following templates computed to second post-Newtonian expansion. They depend, in principle, on several
parameters: the coalescing phase $\phi_c$ and coalescing time $t_c$, and the
parameters corresponding to the physical characteristics of the system, called
intrinsic parameters and globally referred to by the vector $\boldsymbol{\theta}$.
A template is precisely identified by $u(t;\boldsymbol{\theta},\phi_c , t_c)$.  It
is believed that the most relevant intrinsic parameters are the masses of the
binary systems, so as a first approximation it is usual to neglect all intrinsic
parameters except masses. In this approximation, $\boldsymbol{\theta}$ is a vector
of two components.

In a matched filter the signal to noise ratio (SNR) is usually defined by     
\begin{equation}\label{S/N}
\frac{S}{N} \equiv \frac{ \langle h , u \rangle }{\textrm{rms}\langle n , u \rangle }
\end{equation}
where $\langle \dots \rangle$ is a particular inner product defined as:  
\begin{equation}\label{prod}
\langle h , u \rangle \equiv \int_{- \infty}^{+ \infty} \frac{ \tilde{h}^{\ast}(f) \cdot \tilde{u}(f)}{S_{n}(f)}\ df 
\end{equation}

It can be shown that $\textrm{rms}\langle n , u \rangle = (\langle u, u
\rangle)^{1/2}$, so (\ref{S/N}) simplify to $S/N = \langle h,u \rangle $, if
normalized templates are used \cite{flanagan}.

Filtering a signal means to look for local maxima of the signal to noise ratio, in terms of its continuous parameters.  The maximization over the phase
$\phi_c$ can be done analytically ( it can be seen that the maximum value is
obtained computing two inner product as in (\ref{prod}) on two real templates with
opposite phases and then summing their square values \cite{vicere}). Maximization
over $t_c$ instead is achieved at low computational cost calculating the cross
correlations by the FFT algorithm. Maximizations over the intrinsic parameters are
not possible analytically.  For this reason the normal procedure consists in a
discretizations of templates in the space of the intrinsic parameters.

The obvious question concerns the number of templates needed to cover the whole
parameter space. A differential geometrical approach has been developed recently \cite{owenold}.  One introduces a new function, the
\textit{match} $M(\boldsymbol{\theta}_{1},\boldsymbol{\theta}_{2})$, which is the product of two templates with different intrinsic
parameters, where a maximization is assumed over $t_c$ and $\phi_c$: 

%\begin{equation}
%\max_{\Phi_c , t_c } \frac{ \langle h, u(\boldsymbol{\theta},\phi_c , t_c) \rangle }{ \textrm{rms} \langle n, u(\boldsymbol{\theta},\phi_c , t_c) \rangle }
%\end{equation}
\begin{equation}
M(\boldsymbol{\theta}_{1},\boldsymbol{\theta}_{2}) \equiv
\max_{\Delta \phi_c ,\Delta t_c } \langle u(\boldsymbol{\theta}_{1},
\phi_c + \Delta \phi_c, t_c+ \Delta t_c), u(\boldsymbol{\theta}_{2},\phi_c , t_c) \rangle 
\end{equation}

The match between two templates with near equal parameters may be Taylor expanded 
\begin{equation}
M(\boldsymbol{\theta},\boldsymbol{\theta} + \Delta \boldsymbol{\theta}) \simeq 1 + \frac{1}{2} \left( \frac{\partial^{2} M(\boldsymbol{\theta}, \boldsymbol{\omega})}{\partial \omega^{i} \ \partial \omega^{j} } \right)_{\omega^{k} = \theta^{k}} \Delta \theta^{i} \Delta \theta^{j} 
\end{equation}
suggesting the definition of a metric  
\begin{equation}
g_{ij}(\boldsymbol{\theta}) \equiv - \frac{1}{2} \left( \frac{\partial^{2} M(\boldsymbol{\theta},\boldsymbol{ \omega})}{\partial \omega^{i} \ \partial \omega^{j} } \right)_{\omega^{k} = \theta^{k}}
\end{equation}
\begin{equation}
M(\boldsymbol{\theta},\boldsymbol{\theta} + \Delta \boldsymbol{\theta}) \simeq 1 - g_{ij}\ \Delta \theta^{i} \Delta \theta^{j} 
\end{equation}

In the limit of close template spacing we have an analytical function able to
measure the distance between templates in the intrinsic parameter space. The
metric $g_{ij}(\boldsymbol{\theta})$ depends on the intrinsic parameters so the
real volume covered by a template varies locally. This effect can be reduced
writing the templates in terms of some new variables for which 
the metric is more regular. One suitable choice is the following:
\begin{equation}\label{2theta0}
\theta_{1} = \frac{5}{128} \ (\pi f_{0})^{-5/3}\cdot \frac{ M^{-2/3} }{\mu} 
\end{equation}
\begin{equation}\label{2theta1}
\theta_{2} = \frac{1}{4} \ \pi^{1/3} f_{0}^{-5/3}\cdot \frac{ M^{1/3}}{ \mu}
\end{equation}
where $M$ is the total mass of the binary system, $\mu$ the reduced mass, and
$f_{0}$ an arbitrary frequency. This change of variables makes the
metric tensor components constant at the first post-Newtonian order,
so only small $\boldsymbol{\theta}$-dependent contributions are present at the second order approximation.

It is now possible to simply estimate the total number of templates necessary
to recover the signal at a given level of accuracy. We calculate
the volume covered by a single template in the parameter space in term of a
minimal value for the match, the so called \textit{minimal match} $MM$ which
states a minimal requirement on signal recovering capabilities.  For example,
if we simply use a face centered hyper-cubic lattice, we can write the maximum
covering volume with: 
\begin{equation} \label{volume}
\Delta V = \left( 2 \sqrt{ \frac{(1 - MM)}{D}} \right)^{D}
\end{equation}
where $D$ is the dimension of the parameter space (2 in our example).

An approximate estimation of the total template number, applicable when
$N$ is very large \cite{owen}, is given taking the ratio between the total
volume of the physically relevant parameter space and the volume covered by a
template placed in the center of a lattice tile  
\begin{equation}\label{numtot}
N(MM) = \frac{ \int d^{N}\theta \ \sqrt{\textrm{det} g} }{ \Delta V } 
\end{equation}
Using (\ref{numtot}) we estimate that in the range from $0.25$ to $10.$ solar
masses the total template number is roughly $3.8\cdot 10^4$ for LIGO
and $1.1\cdot 10^6$ for VIRGO ( see Section 5 for the additional assumptions
involved in this calculation ).\\  

A last remark we want to make is that the minimal match requirement also
determines a threshold value for the signal ( and templates ) sampling
frequency. This frequency can be simply estimated and will be take into account
later on in our computational estimates.

\section{General Strategy}

In this section we present some observations about a general strategy to compute
correlations. Here we consider an \textit{ideal} case in which most
computer-related issues are neglected. We also limit our treatment only to the
\textit{stored templates strategy}, where templates are pre-calculated, then
Fourier transformed and prepared to be processed and finally stored in memory.
This ideal case is not unrealistic, given the pace at which actual memory sizes
increase in real computers. The quantity to be evaluated
on every template is given by

\begin{equation}\label{correl}
C(t) = \left| \ \int_{- \infty}^{+ \infty}  \tilde{u}^{\ast}(f) \cdot \tilde{s}(f)\ e^{- i 2 \pi f t}\ df \ \right| \ , \ \tilde{u}(f) = \frac{\tilde{h}(f)}{ S_{n}(f)}
\end{equation} 
where $\tilde{h}(f)$ is the Fourier transform of a complex template $h(t)$.\\

At present the best way of compute $C(t)$ uses a FFT algorithm, reducing the
number of needed operations from  $n^{2}$ to $n \log_{2}n$. The FFT algorithm
assumes input periodicity, while in our case signal and templates are not
repeated data.  The usual trick to overcome this problem \cite{numrecipe}
consists in  \textit{padding} with a certain number of zeros the tail of the
templates to be processed. Assume that the template has $n_T$ points. We pad it
so its total length become $n_P$, and then compute the correlation by using the
padded template and $n_P$ signal points. The resulting correlations are only
valid in their first $n_P - n_T$ points, all remaining points being affected by
the periodicity assumption implied in the FFT technique. We define
padding-ratio the quantity $R_{pad} = n_P/n_T$. The result obtained in this way
covers a time-period of length $(n_P - n_T)/f_s$, where $f_s$ is the sampling
frequency of the experimental signal. The last $n_T$ data-points will have to
be re-analyzed in a successive analysis.

The computing power necessary for an on-line analysis of templates of given $n_T$
and $n_P$ (floating point operations per second) is given by:

\begin{equation}\label{costAB}
C_{p} = \frac{f_{s}}{n_{P} - n_{T}}\cdot \left( A \cdot n_P \log_{2}(n_P) + B \cdot n_P \right)
\end{equation}

$A$ and $B$ are constants, usually of the same order, depending on the specific
algorithm used.  In this paper we use a simple-minded FFT algorithm for power of two
length vectors that involves $A=5$ and $B=12$ for the whole analysis. Although
more general and efficient algorithms exist, our choice does not influences
strongly the following observations and final results.

\begin{figure}[htb] 
\epsfig{file=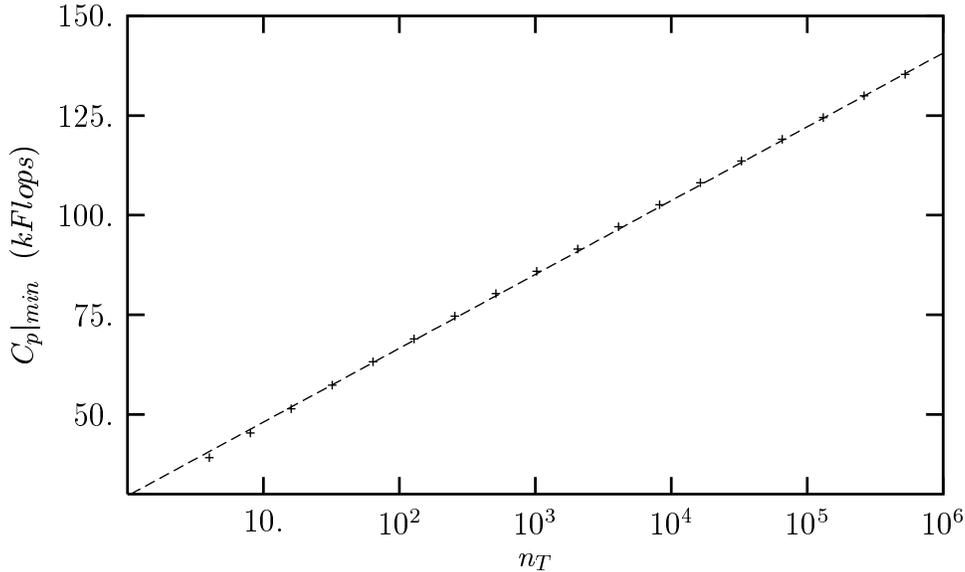,width=\hsize}
\caption{\label{minimum} Estimate of the computing power (floating point
operations per second $Flops$), versus template length $n_{T}$ 
for an optimal choice of $n_P$. We set  $A=5$ e $B=12$ and $f_s = 1 kHz$
(see Eq. \ref{costAB})}
\end{figure}

\begin{figure}[htb] 
\epsfig{file=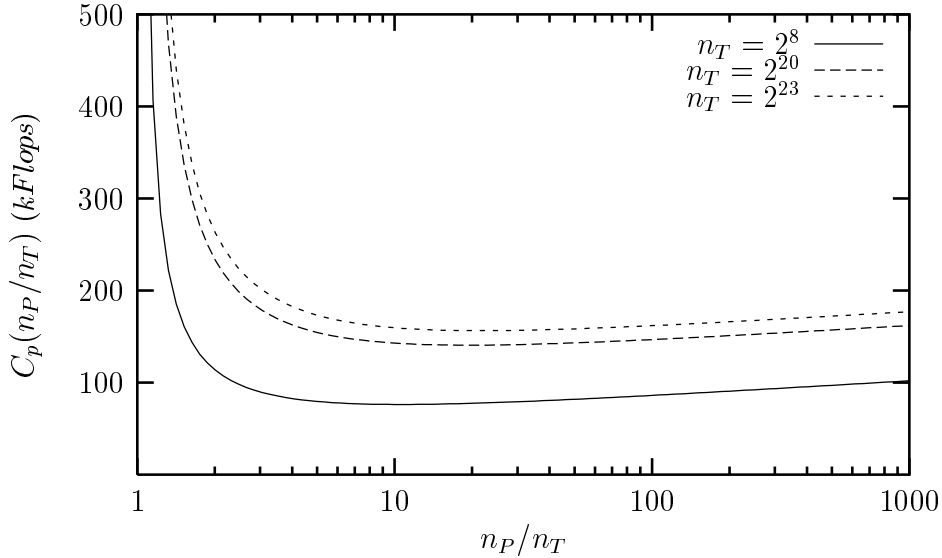,width=\hsize}
\caption{\label{best} Computing power in $Flops$ versus the padding ratio
$n_{P}/n_{T}$, for three typical template lengths. We use the same parameters
as in the previous figure.}
\end{figure}

One interesting question
concerns the optimal padding that minimizes computing requirements. If one
disregards the fact that (\ref{costAB}) holds only for $n_{P}$ values that are
powers of 2, the answer is given by fig.\ref{minimum},  where the minimum in $n_P$
of eq. 14 is plotted as a function of $n_T$, for $f_{s}=1 kHz$. The behavior is
very close to a logarithmic function in $n_T$, so computing costs depend very weakly on $n_T$.
This result is obtained for an optimal choice of $n_{P}$, as discussed above. As shown in
fig.\ref{best}, the optimal value for $n_P$ grows with $n_T$, implying in
principle very large memory requests. In practice however (see again
fig.\ref{best}) a value of $n_P/n_T \simeq 2\cdots 4$ is very close to the optimal
case for reasonable values of $n_T$.  This finally means that deviations from the
optimal padding length do not produce drastic consequences on the computing
power needed to perform the analysis, and that $n_{p}$ can be easily adjusted to a
suitable power of two.  

\section{Analysis on APEmille}\label{ape}

The APE family of massively parallel processor has been developed in order to
satisfy the number crunching requirements of Lattice Gauge Theories (LGT)\cite{ape}.
Machines of the present APE generation (APEmille) are installed at several
sites, delivering an overall peak processing power of about 2 Tflops. The
largest sites have typically 1000 processing nodes (i.e., 520 Gflops)
\cite{apeberlin}. Sustained performance on production-grade LGT codes is about
45 \% of peak performance. A new APE generation (APEnext) is under development,
and expected to reach the physics-production stage in early 2004. $O(10Tflops)$ 
peak performance installations are being considered.

APEmille systems are based on a building block containing 8 processing nodes
(processor and memory) running in Single Instruction Multiple Data (SIMD)
mode. Each processor is optimized for floating point arithmetics and
has a peak performance of 500 MFlops in IEEE single precision mode.
The processors are logically assembled as the sites of a 
$2 \times 2 \times 2$ mesh, with data links connecting the edges. This
arrangement is called a ``cluster'' or a ``Cube''. 

Large APEmille systems are based on a larger 3-dimensional mesh of processor,
based on replicas of the above-described building block. The resulting mesh
has a full set of first neighbor communication links. In a typical LGT
application the whole system works in lock-step mode as a single SIMD system.
More important for the present application, each Cube is able to operate
independently, running its own program under the control of a Linux-based
personal-computer acting as a host. There is one host machine every 4 Cubes. A
set of up to 32 Cubes (i.e., 256 nodes) and the corresponding 8 host machines
is a fully independent unit housed in a standard-size mechanical enclosure. 
Each Cube has access to networked disks with a bandwidth of about 4 MByte/sec.
In some APEmille installations, disks have been mounted directly on the host
PCs. In this case, bandwidth increases approximately by a factor 4.

The next generation APE system (apeNEXT) is, for the purposes of the
present discussion, just a faster version of the same architecture. The only
(welcome) architectural difference is the fact that the basic logical
building block (capable of independent operation) is now just one processing node.

A large APEmille system can be seen as a large farm of processors, whose basic
element is a SIMD machine of dimension 8. A better way to look at the
SIMD cluster in our case follows the paradigm of vector computing: the SIMD
cluster applies the input signal to a vector of 8 templates and produces a
vector of 8 correlations. In a variation of the same method,
the same template could be present on all nodes of the SIMD cluster, and
correlations at 8 staggered time points could be computed.
Since the number of correlations is of the order of
$10^5 \cdots 10^6$, each element of a large farm (say $10^3$ SIMD clusters)
takes responsibility for several hundreds or thousands of templates. This
is good news, since APE processors can exploit vector processing within the
node to reach high efficiency (we just recall here for reader interested in
architectural details that vector processing effectively helps to hide memory
access latencies).

We have written an APE code performing all the steps needed for matched filtering on
a pre-calculated (and pre-FFT transformed) set (vector) of $k$ templates
each of length $n$, and measured its performance on
an APE cluster. An analysis of the details of the APEmille processor
suggest to model the computation time $T_C(n,k)$  as

\begin{equation}\label{costc}
T_{C}(n,k) = f(n) \cdot g(k) \cdot k   
\end{equation}

\begin{figure}[htb]
\epsfig{file=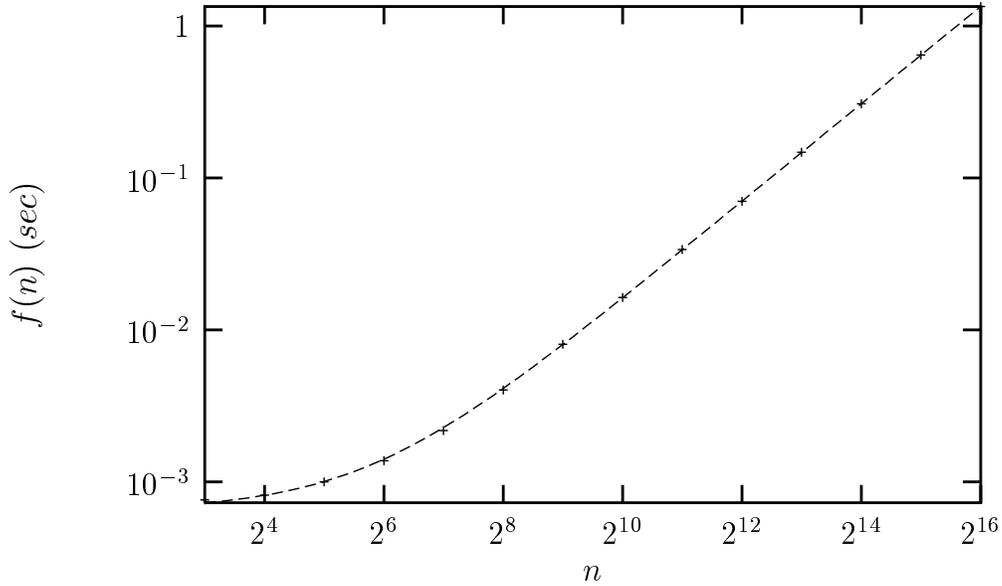,width=\hsize}
\caption{\label{f_n} Analysis time as a function of template length
$T_{C}(n,1) = f(n)$. Measured data points are fitted to the functional form
of \ref{costc}, with $c_1 = 8.6 \cdot 10^{-7} sec$,
$c_2 = 6.6 \cdot 10^{-6}sec $ and $c_3 = 6.6 \cdot 10^{-4} sec$.}
\end{figure}

\begin{figure}[htb]
\epsfig{file=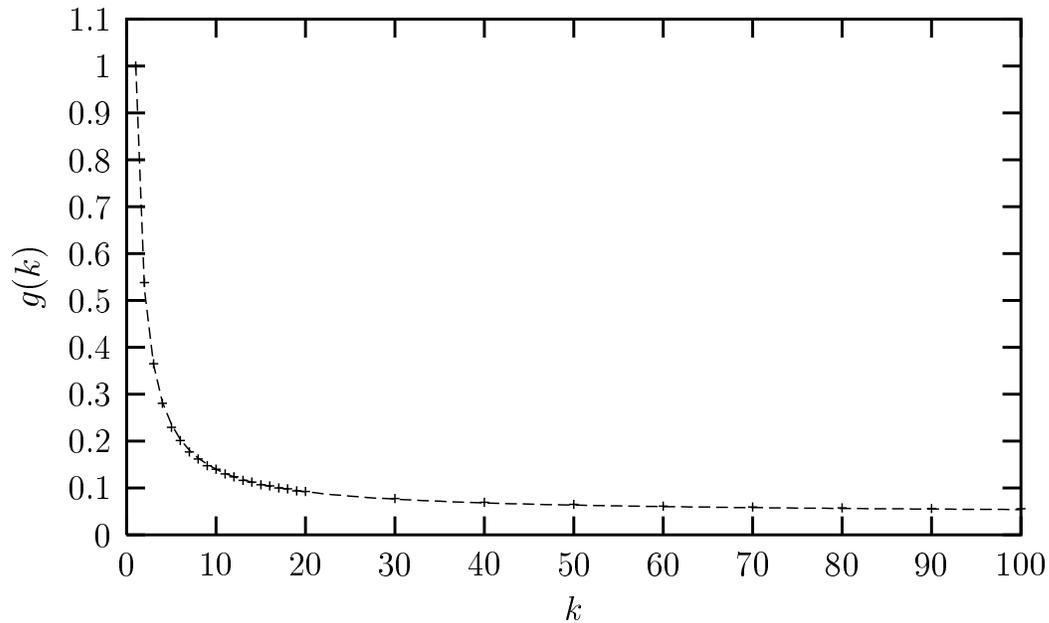,width=\hsize}
\caption{\label{g_k} Normalized processor efficiency $g(k)$
as a function of the vector length $k$. Measured points are fitted to
(\ref{gk}).}
\end{figure}

$f(n)$  is related to the complexity of the computation, that we
model as $c_1 n \cdot \log_{2}(n) + c_2 n + c_3$, following Eq.\ref{costAB}
and introducing one more parameter ($c_3$) covering machine effects. $g(k)$ is
a measure of the processor efficiency as a function of the vector length $k$,
that we normalize to $g(1)=1$.
Taking into account that the computation is memory-bandwidth limited (as
opposed to processing-power limited), we adopt the following functional form
for $g(k)$:

\begin{equation}\label{gk}
g(k) = c_4/k + c_5
\end{equation}
Measured and fitted values for
$f(n)$ and $g(k)$ are shown in fig.\ref{f_n} and fig.\ref{g_k} respectively.


APEmille efficiencies are smooth functions of $n_P$ and $k$.
A rather good value of $\simeq 20\%$, including all computational overheads,
is possible when large sets of templates ($k \ge 20$) are used. 
\section{Allocation criteria and processor numbers}\label{ourcriteria}

A general templates allocation strategy on real computers has to take into
account the limited size in memory and the available computing power available. 
Here we present some quantitative aspects of memory and CPU usage involved in
our analysis, then we give our allocation criteria for the optimal template
number manageable by a single processor.

This discussion focuses on criteria that are appropriate for the APE family of
processors. The focus is to exploit vectorization as much as possible
and to find ways to reduce input-output bandwidth requirements,
so 
our discussion can be applied to a larger class
of processors.

We start from memory. Each processor has $k$ stored templates of similar
length $n_{T}$. (In the APEmille case, the term processor must be understood
to refer to the basic cluster of 8 processing element).

Vector processing of all the templates requires that they are all padded to the
same $n_{P}$, so we need $k$ arrays of $n_P$ complex words, and
matching space for the final correlation results.

There are two basic memory allocation strategies: we may assign different sets
of $k$ templates to each element in a basic 8 processor cluster, and have all
of them compute the corresponding correlations for the same time stretch
$(n_{P} - n_{T}) / f_s$, so each cluster computes $8 k$ correlations.
Alternatively, we may assign the same set of templates to all processing
elements and have each of them compute correlations for different time
intervals. With this choice $k$ correlations are computed for a longer time
stretch $8 (n_{P} - n_{T}) / f_s$.  The best choice between these two nearly
equivalent cases is based on bandwidth constraints. In APEmille, data items
reaching the cluster can be delivered to just one element, or broadcast to all
of them. In the latter case, bandwidth is effectively multiplied by a large
factor ($\times 8$), so there is an advantage if large data blocks must be
broadcast to the complete cluster. We will use quantitatively these
observations later on in this section.

We now consider processing power. The real-time requirement stipulates that
each processor cluster completes processing all its templates within an
elapsed time $(n_{P} - n_{T}) / f_s$  (or $8 (n_{P} - n_{T}) / f_s$). As shown later
on, for several realistic templates sizes, the processing time $T(n_{P},k)$ is
much shorter that the elapsed time for the $k$ value allowed by memory
constraints. We may therefore try to use the same cluster for a different set
of templates. This may become inefficient since loading a large data base (the
new set of templates) may be a lengthy procedure. This cost may be reduced by
using the same templates several times (corresponding to longer elapsed times)
before loading a new set of templates.

We disregard the overhead associated to the output of the computer correlations
, that can be made very small taking into account the Gaussian character of the noise ( e.g. a $3 \sigma$-cut could reduce the number of the output correlations to the order of $1\%$ ). More interestingly a cross correlation among closely spaced templates could be performed on line packing more densily the available information. 

We would like to optimize among these conflicting requirements. Let us
consider the total compute time both for different sets of templates (case 1)
or the same set of templates (case 2) on each cluster element.
We have
\begin{itemize}
\item{Case 1:} We want to compute $r$ sets of $8k$ correlations each on
templates of length $n_{P}$, corresponding to the same time interval. We compute
correlations on $l$ adjoining time intervals before switching to a new
set of templates. The computation time can be modeled as
\begin{equation}\label{cost1}
8 r k n_{P} / B + n_{P} l / B + l r T_c(n_{P},k)
\end{equation}

where B is the cluster input-output bandwidth (measured in words per unit time).
The first term in  (\ref{cost1}) is the time required to load the templates on all processors,
the second term is the time needed to broadcast $n_{P}$ signal points to all cluster
elements while the third term refers to the actual computation, to be performed $l
r$ times. Templates, correlations and input data must fit inside the memory,
implying that $(2k + l)n_{P} \le M_T$, where $M_{T}$  is the available memory on each node ( measured in units of complex words ). Also, the computation must complete in a time
interval $ l (n_{P} - n_{T}) / f_s$. In (\ref{cost1}) we assume that all data-points
are loaded once. This reduces input-output time but reserves a large fraction of
memory space to data-points (as opposed to templates). Alternatively (case 1b), we
may load a smaller set of data-points every time we start a new computation. The
corresponding compute time becomes

\begin{equation}\label{cost1b}
8 r k n_{P} / B + n_{P} r l / B + l r T_c(n_{P},k)
\end{equation}

while the memory constraint changes to $(2k + 1)n_{P} \le
M_T$. For any physical template of length $n_{T}$, we must maximize
$8 r k$ in terms of $r$, $k$, $l$ and $n_{P}$ satisfying all constraints.
\item{Case 2:} The procedure discussed above can be applied also in this case. The
corresponding processing time is given by
\begin{equation}\label{cost2}
r k n_{P} / B + 8 n_{P} l / B + l  r T_c(n_{P},k)
\end{equation}
This equation differs from (\ref{cost1}) since we now broadcast
templates while we load different data-points to each processing elements. The
memory constraint is  the same as in Case 1,
while the maximum allowed processing time is $8 l(n_{P} - n_{T}) / f_s$.
Case 2b (multiple data loads) is also easily computed as
\begin{equation}
r k n_{P} /B + 8 n_{P} r l / B + l r T_c(n_{P},k)
\end{equation}
In case 2, we are interested in optimizing $rk$ in terms of the same parameters
as in the previous case.
\end{itemize}

There is one free parameter in the optimization process ($l$). If we increase
$l$ we reduce the relative cost associated with template loading, but increase
the latency associated to the computation. We arbitrarily decide to keep $l$
small enough so the latency for any $n_{T}$ is not longer that a fixed amount
of time $T_W$. We choose $T_W$ as the time length of the longest template
contained in the set. This choice may be useful also for data-organization
purposes: every $T_W$ time interval all correlations corresponding to templates
of all lengths are made available. The result of the optimization process are
given in Table 1 for APEmille and Table 2 for apeNEXT. Results depend weakly on
the allocation procedure discussed above, and are largely dominated by the
sustained processing power. Bandwidth limitations are neatly dealt with: if we
increase the available bandwidth by a factor four (e.g., using local disks) 
the number of templates handled by each cluster increases by less than 10\%.
With our choice of parameters Case 1b is the preferred one for almost all
template lengths.\\
\vspace{1.cm}
\begin{table}[htb] 
\begin{center} 
\begin{tabular}{|l|r|r|r|r|}
\hline
$n_{T}$  & Case 1  & Case 1b& Case 2 & Case 2b \\
\hline
\hline
%        &    3    &   4    &    1   &   2  \\
$2^{8}$  &   4824  &  $\mathbf{4955}$  &   4937 	&  4854\\
$2^{9}$  &   4344  &  $\mathbf{4549}$  &   4535 	&  4382\\
$2^{10}$ &   3720  &  $\mathbf{4016}$  &   4003 	&  3775\\
$2^{11}$ &   3177  &  $\mathbf{3474}$  &   3458 	&  3252\\
$2^{12}$ &   2511  &  $\mathbf{2904}$  &   2885 	&  2606\\
$2^{13}$ &   1792  &  $\mathbf{2226}$  &   2197 	&  1893\\
$2^{14}$ &   1143  &  $\mathbf{1558}$  &   1526 	&  1315\\
$2^{15}$ &    657  &  $\mathbf{1091}$  &   1057 	&   865\\
$2^{16}$ &    349  &   $\mathbf{679}$  &    644 	&   510\\
$2^{17}$ &    180  &   $\mathbf{378}$  &    329 	&   274\\
$2^{18}$ &      -  &   191    	       & $\mathbf{201}$ &   135\\
$2^{19}$ &      -  &    $\mathbf{86}$  &	 -   	&    - \\
\hline
\end{tabular}
\end{center}
\caption{Number of templates handled by each APEmille processor cluster,
as a function of the template length $n_{T}$. Parameters are (see the text
for definitions) $f_s = 2048 Hz$, $B = 5 \cdot 10^{5} W_{c}/sec$, $T_W = 1024 sec$.
Numbers in bold flag the best case , while $-$ mark cases where allocation can not be performed due to memory limits.}
\label{tabmille}
\end{table}                            
%\vspace{.5cm}

We remark that processing time has been directly measured on APEmille,
while apeNEXT values are extrapolations obtained by appropriately
re-scaling the basic machine parameters, such as memory size, I/O bandwidth, and
processor frequency. 

\begin{table}[htb] 
\begin{center} 
\begin{tabular}{|l|r|r|r|r|}
\hline
$n_{T}$  & Case 1  & Case 1b& Case 2 & Case 2b \\
\hline
\hline
%        &    3    &   4               &    1   	&   2  \\
$2^{8}$  &   15631 &  $\mathbf{15765}$ &  15750 	&  15704\\
$2^{9}$  &   14599 &  $\mathbf{14834}$ &  14815 	&  14728\\
$2^{10}$ &   13626 &  $\mathbf{13852}$ &  13832 	&  13752\\
$2^{11}$ &   12455 &  $\mathbf{12844}$ &  12812 	&  12673\\
$2^{12}$ &   10969 &  $\mathbf{11597}$ &  11546 	&  11321\\
$2^{13}$ &   9401  &  $\mathbf{9992}$  &   9917 	&  9735\\
$2^{14}$ &   7782  &  $\mathbf{8614}$  &   8528 	&  8264\\
$2^{15}$ &   5881  &  $\mathbf{6907}$  &   6790 	&  6466\\
$2^{16}$ &   3979  &  $\mathbf{4991}$  &   4852 	&  4540\\
$2^{17}$ &   2409  &  $\mathbf{3418}$  &   3280 	&  3102\\
$2^{18}$ &   1282  &  $\mathbf{2205}$  &   2107         &  1944\\
$2^{19}$ &    673  &  $\mathbf{1259}$  &	1189    &  1087\\
$2^{20}$ &      -  &   641             & $\mathbf{657}$ &   546\\
$2^{21}$ &      -  &   $\mathbf{285}$  &	   -    &     -\\
\hline
\end{tabular}
\end{center}
\caption{Same as \ref{tabmille}, extrapolated to apeNEXT. We re-scale processing
power per node by a factor 3 and bandwidth by a factor 4. Memory increases by
a factor 4 and $T_W = 4096 sec$..}
\label{tabnext}
\end{table}                            
%\vspace{.5cm}

\section{Actual estimates on processors numbers}

We present here an accurate calculation of the total computational
cost, and thus of the total processors number, required in order to analyze
systems of coalescing binaries whose masses are in a certain range.

The main point of this calculation is the production, given a definite mass
range, of a suitable set of templates covering the corresponding parameter
space. We remark that for our purposes (i.e. an estimation of the computational
cost) we only need a realistic template distribution in the parameter space and
not a precise covering procedure for placing all the templates. Therefore we
will adopt a simplified placing algorithm based on a weighted random generation
method. It is also important to note that the number of templates is very sensible to the shape of
the experimental noise spectrum.

In our template distribution strategy we adopt the regular metric $g_{ij}$
obtained using the variables defined in (\ref{2theta0}),(\ref{2theta1}).  We envisage
a square
lattice on the $\theta_1,\theta_2$ parameter space whose links  $\Delta \theta$
are set to $\Delta \theta = (\min_{\boldsymbol{\theta}}{\sqrt{\textrm{det}g
}})^{-1/2}$, where minimization is on the domain $\boldsymbol{\theta}$ corresponding
to the physical parameters we are interested in.
Under these assumptions the surface
corresponding to a square lattice in the $\theta_1,\theta_2$ space can be
approximated with $S(\boldsymbol{\theta}) = \sqrt{\textrm{det} g }\cdot (\Delta
\theta)^2 $, where $\sqrt{\textrm{det} g }$ is the value calculated in a point at
the center of the square. Hence the minimal surface corresponding to  a lattice
tile is unit.  Now we observe that $S$ is proportional to the number
of templates needed in that square region of the parameter space, and that it is
roughly equal to the number of templates when divided by $\Delta V$ of eq.
(\ref{volume}): 

\begin{equation}
N(\boldsymbol{\theta})_{\textrm{per\ square}} = \sqrt{\textrm{det} g(\boldsymbol{\theta}) }\cdot (\Delta \theta)^2 / \Delta V .
\end{equation}

Finally we allocate to every square in the lattice a number of templates equal
to the rounded value of the previous expression, placing the first one in its
center and the others randomly distributed inside the same region.\\

The noise spectral density $S_{n}(f)$, an experimentally measured and fitted
curve, has different behavior for each experiment. It imposes particular
constraints on lower and upper frequency cutoffs and on sampling (or
interpolation) frequency.  In tab.\ref{noise} we list the fitting functions
relative to the VIRGO and LIGO experiment and corresponding parameters that we
use in our calculations.   
\begin{table}[H]
\begin{center}
\begin{tabular}{|c|c|c|c|c|c|}
\hline 
\multicolumn{6}{|c|}{$S_n(f) = S_{p}f^{-5} + S_{m}f^{-1} + S_{s}\left( 1 + \left(f/f_{knee}\right)^2 \right)$}\\
\hline
\hline
Experiment&
\( f_{\mathrm{seism}} \)&
\( S_{\mathrm{p}} \)&
\( S_{\mathrm{m}} \)&
\( S_{\mathrm{s}} \)&
\( f_{\mathrm{knee}} \)\\
\hline 
%GEO600&
%\( 50 \)&
%$4.1\cdot10^{36}$&
%$9.\cdot10^{43}$&
%$1.\cdot10^{44}$&
%577\\
%\hline 
%LIGO 2K&
%\( 40 \)&
%$2.1\cdot10^{35}$&
%$2.25\cdot10^{43}$&
%$4.35\cdot10^{46}$&
%182\\
%\hline 
LIGO 4K&
\( 40 \)&
$5.6\cdot10^{36}$&
$3.9\cdot10^{44}$&
$1.1\cdot10^{46}$&
83\\
\hline 
%TAMA&
%\( 50 \)&
%$6.6\cdot10^{31}$&
%$3.2\cdot10^{40}$&
%$1.78\cdot10^{42}$&
%500\\
%\hline 
VIRGO&
\( 4 \)&
$9.\cdot10^{37}$&
$4.5\cdot10^{43}$&
$3.24\cdot 10^{46}$&
500\\
\hline
\end{tabular}
\caption{Spectral density noise parameters for LIGO and VIRGO from \cite{vicere2}.} \label{noise}
\end{center}
\end{table}                            

The noise curves for LIGO and VIRGO are quite different. While LIGO is very
sensible in a narrow frequency interval VIRGO has a lower peak sensibility but
is better in a wider range of frequencies.

In tab.\ref{noise} $f_{seism}$ indicates the so called seismic frequency, i.e.
the frequency below which seismic noise is expected to dominate over all other noise sources. Slightly different definitions for $f_{seism}$ exists, see \cite{owen} and
\cite{vicere2}. Integration below this limit does not contribute significantly
to detectability but is quite expensive in terms of computing power since
it increases the number of templates and the template length.\\  
In our
calculations, as proposed in \cite{vicere}, in order to reduce computing requirements we
adopt a more restrictive frequency range. We fix a lower and an upper frequency
bounds $f_l$ and $f_u$ such that the total SNR recovery is at least of $97\%$ (we assume the
lower SNR loss of $2 \%$ and the upper of $1\%$). 
We note that template
lengths are very sensible to the lower frequency cut-off, as the duration time, which influences linearly the storage requirements
and logarithmically the computation cost, scales as $f_{l}^{-8/3}$.   
Our frequency bounds 
are reported in the first two column of tab.\ref{cuts}.

Another point concerns the Minimal Match. We set  $MM = 0.97$,
which corresponds to an event rate loss of roughly $10 \%$ \cite{owen}.

As discussed at the end of Section 2 the $MM$ level sets not only the
density of templates in the parameter space (by $\Delta V$, see (\ref{volume}) )
but also the signal ( and templates ) sampling frequency. This frequency could
be very height, so in some cases, memory requirements could be severe. 

An alternative analysis strategy consist in sampling the signal at a lower frequency
and then obtaining correlations at half-time points by an interpolation. In fact
this can be simply achieved performing further anti-Fourier transforms after
introducing in the integrals a suitable phase displacement. The number of
interpolations obviously multiplies the analysis time.

In our estimate we fix the sampling frequency by $f_{s} = 2 f_{u}$ (rounding
its value to the greatest power of two), so we have $f_{s} = 1024\ Hz$ for LIGO
and $f_{s} = 2048\ Hz$ for VIRGO. This means ( see last column in
table\ref{cuts}) that we have to compute correlations at intermediate times by
interpolations once for VIRGO and once for LIGO experiments.  

\begin{center}
\begin{table}[H]
\begin{tabular}{|c|c|c|c|c|}
\hline
Experiment & $f_{l}\ (Hz)$ & $f_{u}$ &SNR recovery & $f_{int}$ at $MM = 0.97$ \\
\hline
\hline
LIGO 4K &
55 (2.0 \%) &
390 (1.0 \%) & 
97.0\% &
1203\\
VIRGO &
26 (1.9 \%) & 
900 (1.1 \%)& 
97.0 \%&
3253\\
\hline
\end{tabular}
\caption{Frequency cuts used in our analysis,
depending on signal to noise recovery and on Minimal Match.} \label{cuts}
\end{table}                            
\end{center}  

\begin{figure}[htb] 
\epsfig{file=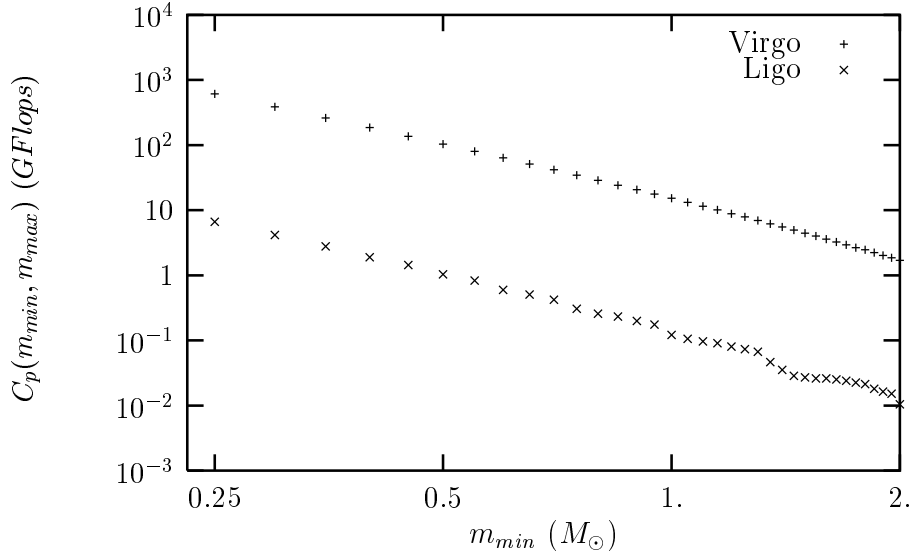,width=\hsize}
\caption{\label{ideal} Behavior of the total computation cost (in GFlops) in the infinite memory availability case versus the lower mass limit, where maximum total mass is $10.\ M_{\odot}$. Here we use VIRGO and LIGO noise spectrum, freq. sampling at 2048 Hz and 1024 Hz respectively and $MM = 97\%$.}
\end{figure}

First, we show in fig.\ref{ideal} the total computational cost to compute the
correlation for binary systems whose masses are in a range of $m_{min}$ to $10$
solar masses, as a function of $m_{min}$, under the assumption of optimal
padding. We use the parameters listed in tables
\ref{noise} and \ref{cuts}.
The computational cost roughly follows a (fitted) power-law behavior, with
exponent of the order of $2.4$. This behavior can be easily guessed, taking
advantage of the fact that the computational load of each template depends very
weakly on its length, and that the $det g(\theta)$ depends weakly on the
$\theta$ variables. Under these assumptions the computational cost scales up to log-corrections as the
area of the region in $\theta$ space corresponding to a given interval of
allowed star-masses. The latter can be easily shown by power counting to behave
as $m_{min}^{7/3}$.\\

The large difference in computational cost between the two experiment, clearly
noticeable in fig.\ref{ideal}, derives, although in a complex way, from the
different noise spectra and from the correspondingly different frequency cuts.   

We now specialize the discussion to APE systems. We proceed
establishing a mass interval, then generating its template distribution.  We
``stretch'' template lengths to the nearest power of two larger  than the
actual length (a slightly pessimistic assumption). Finally we divide each group
of templates of equal length by the corresponding number of templates handled
by one processor cluster ( the bold numbers in tab.1 and tab.2), and sum all
the resulting quotients. The final result represent the number of APE processor
needed to satisfy the real time requirement on the given mass interval. 

\begin{figure}[htb]
%\hskip -.4cm
\epsfig{file=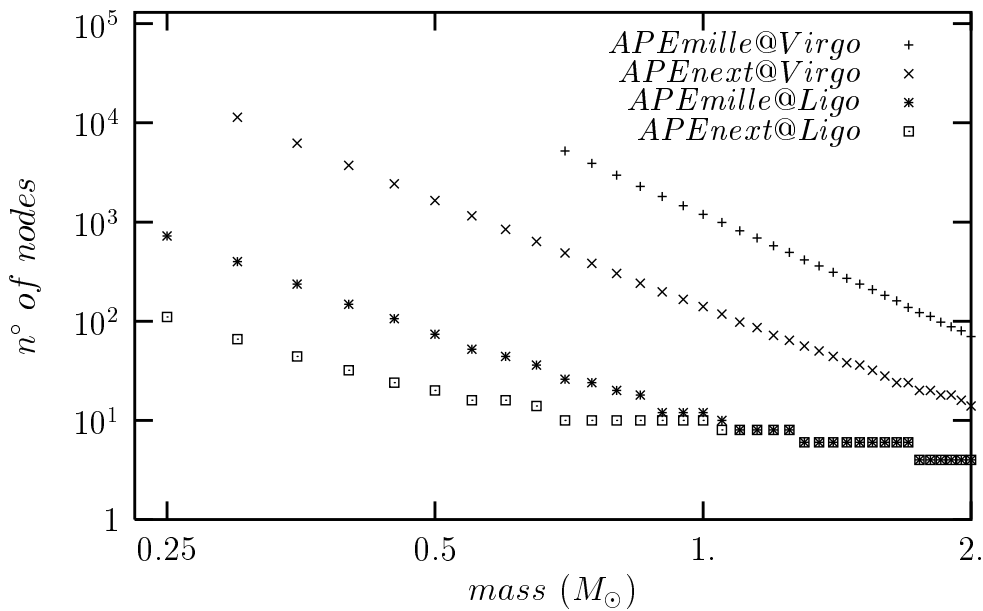,width=\hsize}
\caption{\label{proc} Number of needed APE nodes versus the lower mass limit, maximum total mass is $10.\ M_{\odot}$. Using VIRGO and LIGO noise spectrum, freq. sampling of 2048 Hz and 1024 Hz respectively, $MM = 97\%$ .}
\end{figure}

The computational cost of this matching filter analysis is particularly 
sensible to the lower mass limit because of the increasing template length and
of the irregular behavior of the metric tensor $g_{ij}$ in that region of the
parameter space.  For this reason it is useful to plot the number of
processor versus the lower mass limit.
The number of nodes (one cluster consist of 8 nodes) for a mass interval
from  $m_{min}$ to $10 M_{\odot}$ is plotted in fig. \ref{proc}, where we use
noise spectra relevant for LIGO and VIRGO. This complete our analysis. 

\section{Conclusions}
In this paper we have developed a reliable estimate of the
computational costs for real-time matched filters for GW search from binary
star systems, in a massively parallel processing environment.

We have analyzed some criteria to optimally allocate the processing load to a
farm of processors. We have written a code performing the analysis on an APE
system and we have measured its performances. Our result is  that available
(APEmille) systems are able to satisfy the requirements of a real-time analysis
of the complexity corresponding to the LIGO experiment in the mass range
between 0.25 and 10 $M_{\odot}$.

The VIRGO experiment ( with its lower and wider noise curve )  has
substantially larger computing requirements that cannot be fulfilled by an
APEmille system in the same mass range. The new APE generation, expected to be
available in early 2004, partially closes this performance gap.

\section*{Acknowledgments}
We thank F. Vetrano for reading our manuscript. T. Giorgino and F. Toschi wrote
the FFT code for APEmille. This work was partially supported by Neuricam spa,
through a doctoral grant program with the University of Ferrara.

\end{document}